\documentclass[aps,prl,preprint,superscriptaddress]{revtex4-1}
\usepackage{graphicx}
\usepackage{amssymb}
\usepackage{amsmath}
\usepackage{hyperref}
\usepackage{float}
\usepackage{mathtools}
\usepackage{dcolumn}
\usepackage{xcolor}
\begin{document}

\title{Ultra-high permeable phenine nanotube membranes for water desalination}

\author{Supriyo Naskar}
\affiliation{Center for Condensed Matter Theory, Department of Physics, Indian Institute of Science, Bangalore 560012, India}
\author{Anil Kumar Sahoo}
\affiliation{Center for Condensed Matter Theory, Department of Physics, Indian Institute of Science, Bangalore 560012, India}
\affiliation{Max Planck Institute of Colloids and Interfaces, Am Mühlenberg 1, 14476 Potsdam, Germany}
\affiliation{Fachbereich Physik, Freie Universität Berlin, Arnimallee 14, 14195 Berlin, Germany}
\author{Mohd Moid}
\affiliation{Center for Condensed Matter Theory, Department of Physics, Indian Institute of Science, Bangalore 560012, India}
\author{Prabal K. Maiti}
\email{maiti@iisc.ac.in}
\affiliation{Center for Condensed Matter Theory, Department of Physics, Indian Institute of Science, Bangalore 560012, India}

\begin{abstract}
Nanopore desalination technology hinges on high water-permeable membranes which, at the same time, block ions efficiently. In this study, we consider a recently synthesized [\emph{Science} \textbf{363}, 151–155 (2019)] phenine nanotube (PNT) for water desalination applications. Using both equilibrium and non-equilibrium molecular dynamics simulations, we show that the PNT membrane completely rejects salts, but permeates water at a rate which is an order-of-magnitude higher than that of all the membranes used for water filtration. We provide the microscopic mechanisms of salt rejection and fast water-transport by calculating the free-energy landscapes and electrostatic potential profiles. A collective diffusion model accurately predicts the water permeability obtained from the simulations over a wide range of pressure gradients. We propose a method to calculate the osmotic water permeability from the equilibrium simulation data and find that it is very high for the PNT membrane. These remarkable properties of PNT can be applied in various nanofluidic applications, such as ion-selective channels, ionic transistors, sensing, molecular sieving, and blue energy harvesting.
\end{abstract}

\maketitle

\section{Introduction}
Desalination is one of the most widely used methods for providing fresh water, given the abundance of seawater\cite{2,4,5,6,7}. However, most of these desalination techniques suffer from several significant disadvantages, such as large energy consumption and high capital cost\cite{3,5}. Reverse osmosis(RO)-based desalination is the best energy-efficient technology compared to thermal and other desalination techniques\cite{5,8,9}. In this regard, carbon-based nanoporous materials and several other two-dimensional membranes with fabricated pores show promising properties, where the size of a pore is optimized to allow faster water transport while rejecting ions efficiently\cite{10,11,12,13,14,15,16,17,18,19,yan2017graphene,li2018highly}.\par

Carbon nanotubes (CNT), with the high slip flow enhancement property because of the smooth hydrophobic pore wall, are suited for achieving a higher water permeability\cite{20,21,22,23,24,25,58,59,RN2632,salt,yan2017graphene,li2018highly,joseph2008carbon,majumder2011anomalous}. 
The salt rejection characteristics of CNT membranes can be tuned by the end functionalization with various chemical groups\cite{11,12,16,26,27}; however, the bulk manufacturability of CNT membranes with the precise end functionalization groups is extremely difficult in experiments. On the other hand, 2D materials, such as Graphene, MoS$_2$, etc. with nanopore(s) made on these surfaces and metal-organic frameworks such as Ni/Cu-HAB have been demonstrated in simulations to show excellent water transport rate as well as the percentage of ion rejection\cite{13,14,15,18,19}. However, there is not only the difficulty in making scalable membranes out of these 2D materials for large-scale water filtration but also the complicacy in experiments to fabricate sub-nanometer diameter pores that prevent ion passage.\par

Phenine nanotube (PNT) is a recently synthesized carbon-based nanomaterials\cite{28} which has a similar nanotubular structure as CNT but with periodic defect on the wall (Fig. \ref{fig1})\cite{28}. Interestingly, PNTs self-assemble to form membrane-like structure, and the crystal structure of PNT membrane contains voids that encapsulate fullerene (C60) molecules\cite{28}. In such a membrane, PNTs are found to be highly aligned and packed in a tetragonal geometry, and the membrane has a large void fraction of 63\%\cite{28}. The stable PNT membrane with highly aligned, nanoscale diameter pores inspired us to ascertain its desalination properties.\par

The performance of membrane-based water filtration technology hinges on three parameters, namely, the fraction of salt rejection ($\chi$), the water permeability ($p_w$), and the energy ($E$) required to overcome the osmotic pressure gradient. Therefore, we can define an efficiency parameter $\eta=(\chi \cdot p_w)/E$. 
In this article, we show that all of the above three parameters are optimized for the PNT membrane, making it one of the potential membrane for water desalination application. We simulate PNT of chirality (9,9) with two different end groups, the one is with an experimentally synthesized t-Bu group (referred to as PNT) and the other is with hydrogen (referred to as PNT(H)) (Fig. \ref{fig1}). Using molecular dynamics simulations, we explore the water desalination efficiency of PNT membranes as a function of applied hydrostatic pressure and end-functional groups.  The microscopic mechanism of high water flux and salt rejection for the PNT membrane is explained through free-energy and electrostatic map calculations as well as theoretical modelling.\par
\section{Results and Discussions}
\begin{figure*}[htbp]
 \centering
 \includegraphics[width=1.0\linewidth,keepaspectratio=true]{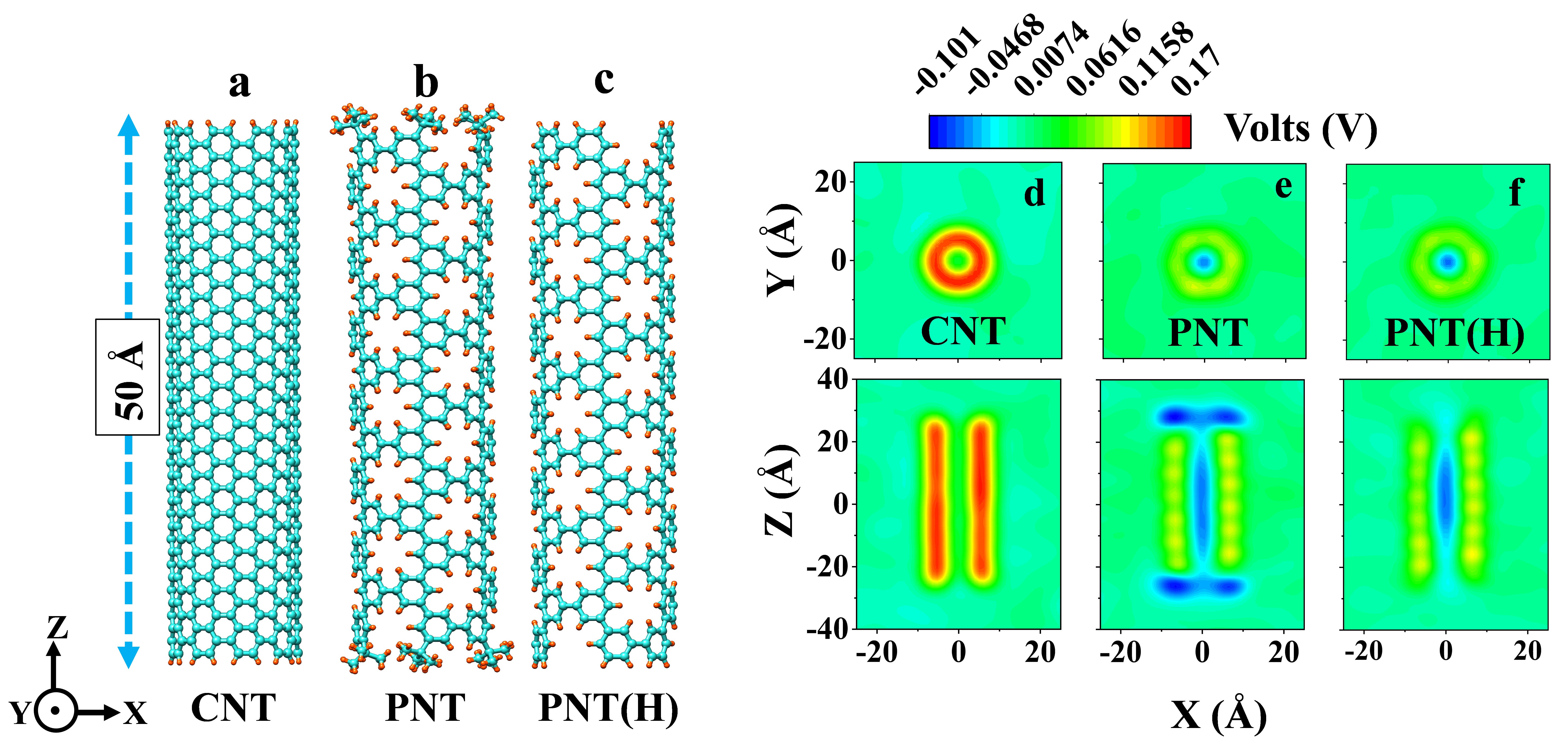}
 \caption{Initial structures, and electrostatics of the PNTs and CNT. In silico model of (a) CNT, (b) PNT with tertiary butyl as end functionalization group and (c) PNT with hydrogen as end functionalization group referred to as PNT(H). The chirality of each of the nanotube is (9,9), and the length is 50 \AA. Cyan represents carbon atoms and red represents hydrogen atoms. (d)-(f) The top view and the side view of the electrostatic potential map of (d) CNT, (e) PNT, and (f) PNT(H) immersed in water.}
 \label{fig1}
\end{figure*}

\begin{figure*}[htbp]
 \centering
 \includegraphics[width=1.0\linewidth,keepaspectratio=true]{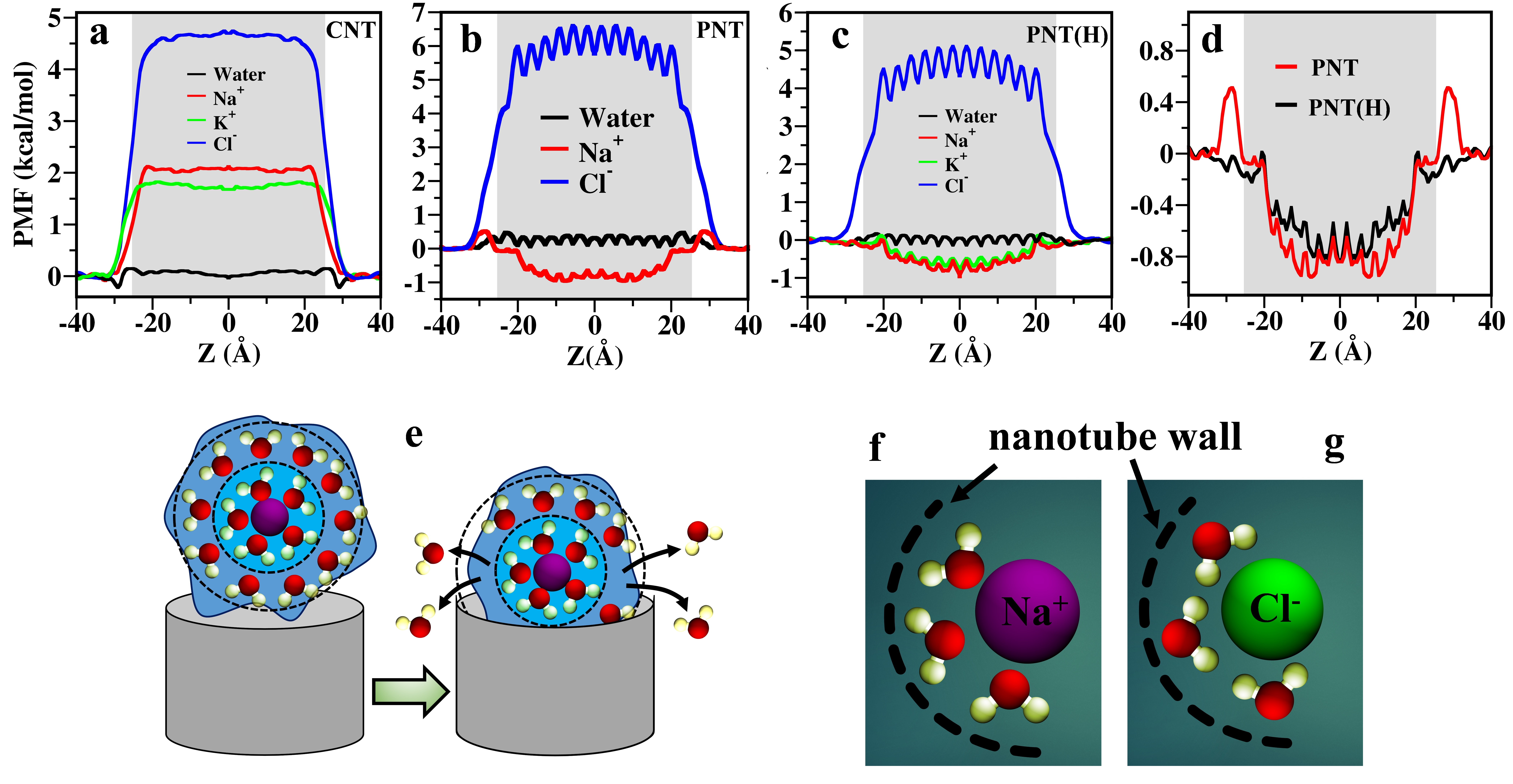}
 \caption{Free energy profiles of water and ions along the axis of (a) CNT and (b) PNT and (c) PNT(H), of chirality (9,9). (d) The free energy profile of Na$^+$ ion for PNT is compared to that of PNT(H). (e) Schematic representation of the molecular mechanism of dehydration of ions while passing through the nanotube. The large hydration shell of ion losses some water as it enters the nanotube. The middle brown circle is the ion and around it, the first and second water solvation shells are shown with different shades of blue color. (f)-(g) Schematic representation of molecular structure of water around the ions inside the nanotube. Purple is for Na$^+$ ion, green is for Cl$^-$ ion, red is for oxygen atom of water and white is for hydrogen atom of water. The more electronegative oxygen atoms of water are pointing towards the Na$^+$ ion; whereas for Cl$^-$, the electropositive hydrogen atoms of water are pointing towards it.}
 \label{fig2}
\end{figure*}

The average electrostatic maps calculated by solving Poisson's equation of a single nanotube solvated in water are shown in Fig. \ref{fig1} (See method section for the details of the calculation). For CNT, the electrostatic potential of the bulk region away from the nanotube and inside the nanotube is almost similar (Fig. \ref{fig1}d). In contrast, compared to the bulk, PNT has a negative electrostatic potential inside it, as well as near its mouth (Fig. \ref{fig1}e). Also, as shown in Fig. \ref{fig1}f, PNT(H) has same electrostatic potential energy profile as PNT, except near its mouth region because of the absence of t-Bu groups. This indicates that PNT or PNT(H) can act as a cation-selective pore and can completely reject anions. Note that CNT rejects ions by the size exclusion principle that includes the solvation energy cost of partial or complete dehydration of the ions (Fig. \ref{fig2}e)\cite{11}. In contrast, for PNT, in addition to the size exclusion principle, the charge expulsion mechanism because of the electrostatic interaction is expected to play an important role in ion rejection.\par
To quantify the dehydration and/or electrostatic effects mentioned above, we compute the free energy of water and ions along the long axis of the nanotube (Fig. \ref{fig2}). For both the CNT and PNT, we find that the water spontaneously enters the pore and move through them without any energy barrier. The periodic oscillation in the free energy profile of water inside PNT (Fig. \ref{fig2}b,c) correctly captures the different energetically-favorable locations of water which are also prominent in the water density profile (See section 3 of SI). In contrast, Cl$^-$ ion faces a free-energy barrier of 5 kcal/mol for entering into CNT; whereas for Cl$^-$ ion entering into PNT, the barrier is slightly higher ($\sim$6 kcal/mol) due to the negative electrostatic potential inside PNT. A completely different behavior is observed for Na$^+$ ion translocation into CNT and PNT. Na$^+$ faces a free energy barrier of $\sim$2 kcal/mol for entering into CNT, whereas for PNT, it faces a much smaller free-energy barrier of $\sim$0.6 kcal/mol to enter inside the pore. Once inside, the free energy decreases to -0.9 kcal/mol inside the pore lumen of PNT (Fig. \ref{fig2}b,c,d). As the CNT pore is in equipotential with the bulk, the barrier arises mainly due to the dehydration of ions (Fig. \ref{fig2}e). In contrast, PNT has a negative potential inside, thus there is a competition between the energy cost due to dehydration of cations and the energy gain due to entering the negative potential region. Compared to the pore lumen of PNT, the dehydration of ions at both the entrances of PNT is more due to the presence of the bulky t-Bu groups. Since there is no such dehydration of ions at the entrances of PNT(H), cations face no such energy barrier (Fig. \ref{fig2}c). The free energy profiles suggest that PNT(H) can be used as a cation selective nanochannel, whereas PNT, which blocks both cations and anions, can be a suitable nanomaterial for water desalination. In order to verify the spontaneous insertion of cations and rejection of anions into PNT(H), we perform eight simulations using both saline water (0.15 M of NaCl or KCl) and seawater (0.6 M of NaCl or KCl) with CNT, PNT(H) (Details of system can be found in table S4 of SI). Indeed, we observe that cations (Na$^+$ and K$^+$) spontaneously move into and translocate through PNT(H), whereas Cl$^-$ ions almost never permeate through the PNT(H) nanotube (section 4 of SI). It is worth mentioning that 0.6 kcal/mol barrier at the entry region of PNT is too small a barrier to hinder the passage of Na$^+$ ions. However, Na$^+$ ions experience a higher barrier for passing through a PNT membrane, as follows. The PMF we calculated is for a single PNT. The PMF do not capture the effect of end-group interaction between two adjacent PNT. Since these end functional groups are bulky in nature, interaction between them leads to reduction of effective radius of PNT in the entry region (Fig. S10 of SI). This results in a much higher barrier for Na$^+$ and Cl$^-$ ions in the entry region for a membrane which cannot be captured in the PMF in a single PNT. \par
Cl$^-$ ions experience a higher free-energy barrier to enter inside both the CNT and PNT, compared to Na$^+$ and K$^+$ ions (Fig. \ref{fig2}a-d). This can also be explained by the dehydration of each ion when it enters into the nanotube, as schematically represented in Fig. \ref{fig2}e. When the ions are in bulk, they are fully hydrated and solvated by water molecules. The size of an ion’s solvation shell depends on its van der Walls (vdW) radius and charge. The vdW radius as well as the radii of the solvation shells of Cl$^-$ ion are larger than that of Na$^+$ and K$^+$ ions. As a result, Cl$^-$ ion has to lose more water molecules compared to the other cations, which causes a higher free-energy penalty for Cl$^-$ (Fig. S5 of SI). Apart from the dehydration effect, additional contributions to the free-energy barrier arise because of the water dipolar effect(Fig. \ref{fig2}f,g). Water orients differently in the vicinities of cations and anions. For a solvated Na$^+$, hydrogen atoms of the water molecules face away from Na$^+$, whereas for Cl$^-$, oxygen atoms of the water molecules face away from Cl$^-$ (Fig. \ref{fig2}f,g). As the electropositive hydrogen atoms of the water molecules like the electronegative carbon atoms of the nanotube wall, there is an extra energy penalty for the entry of a solvated Cl$^-$ ion into the nanopore. In contrast, no extra energy penalty for a solvated Na$^+$ ions. Also, the periodic oscillation in the free energy profiles of the ions inside the PNT occur because when the electropositive hydrogen atoms of the water molecules face the electropositive hydrogen of the PNT wall, the penalty in PMF increases slightly. Similarly, when the electronegative oxygen atoms of the water molecules face the electropositive hydrogen of the PNT wall, the penalty in PMF decreases slightly. Another approach to obtain the free-energy barrier for water and ion transport across the PNT nano-channel is to calculate the Arrhenius activation energy which, however, requires simulations at several temperatures, as done by Babicheva et al. for water desalination by cyclicgraphene and graphenylene \cite{pccp1}. \par

\begin{figure*}[ht!]
 \centering
 \includegraphics[width=0.9\linewidth,keepaspectratio=true]{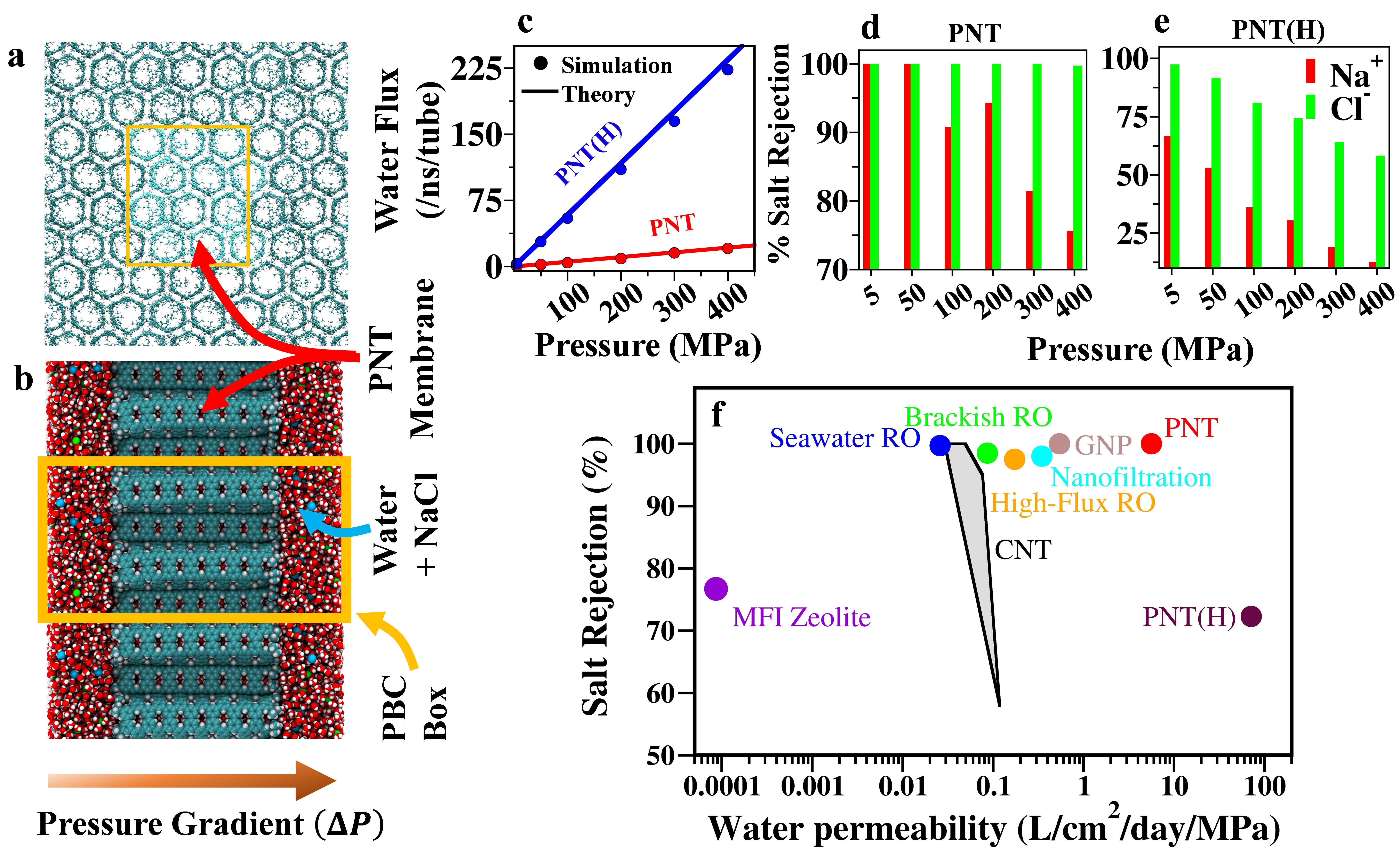}
 \caption{Water permeation and salt rejection through PNT membrane. (a) Initial setup of the membrane where PNTs are arranged in a hexagonal lattice. (b) Side view of the same system with added water and NaCl. The rectangular (orange) box represents the simulation unit shell. (c) Water Flux calculated from non-equilibrium pressure-driven simulations (data points) and from the collective diffusion model (line) for PNT and PNT(H) membranes. The red color is for the PNT membrane while the blue color is for the PNT(H) membrane. Salt rejection efficiency of (d) PNT and (e) PNT(H) membranes. (f) Comparison of water desalination performance of the PNT membrane with existing technologies. The PNT membranes examined in this study permeate water with one-to-several orders of magnitude higher values than the commercial RO membranes\cite{8} with complete salt rejection. Water permeation data for CNT is adapted from Ahn et al.\cite{AHN20121551}. The water permeation data of graphene nanopore (GNP) is taken from Cohen-Tanugi et al.\cite {13} and properly normalized by the area such that the minimum distance between two nanopores is 4 nm, instead of normalizing by only the pore area as done in Refs.\cite{13,cohen2014mechanical}.}
 \label{fig3}
\end{figure*}

We now investigate the effect of hydrostatic pressure gradients on the water flow rate (or flux) and on the percentage (\%) of salt rejection (Fig. \ref{fig3}). We find that the net water-flow rates across both PNT and PNT(H) membranes increase linearly with the applied pressure gradient (Fig. \ref{fig3}c). The water flow rate at each applied pressure is significantly lower for the PNT membrane, compared to the membrane made of PNT(H). This is due to the bulky t-Bu groups present at the entrances of PNT. We also estimate the effective water permeability normalized by the membrane area and applied pressure and compare the performance of PNT membrane with other existing membranes for water filtration in terms of the so-called permeability–selectivity trade-off (Fig. \ref{fig3}f). The water permeability for the PNT(H) membrane is 72 liter/cm$^2$/day/MPa. In contrast, permeability reduces to 6 liter/cm$^2$/day/MPa for the PNT membrane. Most of the commercialized RO membranes like MFI zeolite, Seawater RO, Nanofiltration etc. has a permeability in the range of $10^{-6}$ to $10^{-1}$ liter/cm$^2$/day/MPa. The permeability of PNT membranes is order of magnitude higher than these existing technologies\cite{8}. The water flux through a semi-permeable membrane depends mainly on its porous nature and the amount of void area enclosed as well as the mobility of water inside the nanopore. We find that the water flux through the PNT membrane is very high compared to other conventional membranes. We believe this is mainly due to the presence of large void area of PNT membrane. It is worth mentioning that CNTs having diameters 13.56 Å (corresponding to charility (10,10)) and lower can successfully reject all ions\cite{12,salt}. In contrast, PNTs having a diameter 12.2 Å (corresponding to charility (9,9)) can reject all ions which makes PNT less efficient than CNTs. Several other Carbon nanotube based nanomaterils has also higher permeability than CNTs\cite{yan2017graphene,li2018highly}. Furthermore, in this study, we take the width of the membrane as 5 nm. The water flux can also be increased by reducing the membrane thickness\cite{AHN20121551}. However this affects the mechanical stability of the membrane under an applied pressure. One can, however, achieve a higher mechanical stability by using similar technology employed in conventional RO plants where any water filtration membrane supported by a polysulfone layer\cite{8}. Note that, we have employed in our simulations a wide range of hydrostatic pressures (in the range of 5--400 MPa) which cover the operating pressure of most of the RO plants ($\sim$ 5--10 MPa)\cite{8}. Since the water flow rates for both PNT and PNT(H) membranes increase linearly with the applied pressure, our results also can be extrapolated any pressure regime. It should be noted that energy consumption by the membranes in the RO stage is approaching the thermodynamic minimum. It is pre-treatment, fouling, concentration polarisation etc that are the issues in terms of energy consumption. Novel membrane materials are not likely to reduce the energy consumption of RO. Furthermore, in water desalination, improved cleaning and fouling resistance of membranes are also important aspects to consider. It is impossible to address these issues from numerical calculations or molecular dynamics simulations. Also, we cannot completely rule out the quantum effect on the structure and dynamics of water inside the nanoporins which effects the water permeation rates.
\par 

Another important property of these nanotubes is their capability to block the passage of ions very efficiently. To quantify these ion rejection efficiency, we calculate the ion rejection \%\ for these membranes at different applied pressures. The ion rejection \%\ is calculated using the following definition\cite{salt2}
\begin{equation}
    R(\%)=\ (1-\frac{J_{ion}}{J_{water}}\ \times \frac{C_{water}}{C_{ion}}\ )\times100 
    \label{1}
\end{equation}
where $J$ corresponds to the flux and $C$ corresponds to the concentration (number density) of water or ion. For a membrane to be effective for desalination, the ion rejection should be 100\% with water permeability as high as possible. Our results for the salt rejection suggest that the PNT membrane is an excellent candidate to desalinate water (Fig. \ref{fig3}d). It can reject 100\% Cl$^-$ ion up to 400 MPa and more than 90\% Na$^+$ ion up to 200 MPa. Almost 100\% of all the ions are rejected when the pressure gradient is under 50 MPa. But, the ion rejection \% for the PNT(H) membrane (Fig. \ref{fig3}e) is much less due to the low or negligible free-energy barrier for the ions at the entrance of the nanotubes. Moreover, the ion rejection \% decreases with the increase in pressure. Also, the Na$^+$ ion rejection \% is lower than the Cl$^-$ ion rejection \% due to the nature of the free-energy profiles as discussed earlier. The reduction in the ion rejection \% with the increase in the applied pressure (Fig. \ref{fig3}d,e) can be qualitatively explained by a simple, Arrhenius-like kinetic model\cite{13}. The rate of permeation $dN_i/dt$ of a species, $i$, can be approximated as follows

\begin{equation}
    \frac{dN_i}{dt}=A_i\exp[\beta\Omega_i P-\beta(\Delta E_i-T\Delta S_i)] \label{2}
\end{equation}
where $A_i$ represents the attempt rate of the species $i$, which can be considered constant for a particular membrane at a constant temperature, $\Omega_i$ represents the effective volume of the species $i$, and $P$ is the pressure. $\Delta E_i-T\Delta S_i$ denotes the free-energy barrier faced by the species, $i$, while traversing through the membrane. The salt ions have larger effective volume and face higher free-energy barrier which make their permeation difficult compared to a water molecules (whose free-energy barrier is close to zero). Furthermore, both the effective volume and the free-energy barrier are lower for Na$^+$ ions than Cl$^-$, thus making the permeation of Na$^+$ ions easier. This simple model qualitatively explains the relative difference of permeation rates between different species of ions. Though the model predicts an exponential increase in the permeation rate with the applied pressure, the simulated water-permeation rate, however, increases linearly with pressure (Fig. \ref{fig3}c). We are able to quantitatively reproduce the simulated water-permeation rate versus pressure curves using the collective diffusion model\cite{29}. In this model, the movement of water inside a nanopore is assumed to follow a coupled many-body dynamics and is described by a collective coordinate $n(t)$, which denotes the net amount of water permeation at time $t$. The net water flux $j_w$ can be written as

\begin{equation}
    j_w=\frac{\langle n(t)\rangle}{t}=\frac{fl}{K_B\ T}\ D_n=-\frac{V\Delta P}{NK_B\ T}\ D_n \label{3}
\end{equation}

Here$fl=-Al\Delta P/N= -\Delta PV/N$; $A$ is the cross-sectional area of the membrane, $V (=A\times l)$ is the volume of the simulation box, $N$ is the total number of water molecules, $\Delta P$ is the pressure gradient, and $D_n$ is defined as the collective diffusion coefficient of $n(t)$. The detailed derivation of Eq. \ref{3} is given in section 7 of SI. In equilibrium, the mean square fluctuation of $n(t)$ follows Einstein relation
\begin{equation}
    \langle n^2(t)\rangle=2\ D_n\ t \label{4}
\end{equation}
Following Eq. \ref{4}, we evaluate $D_n$ from the slope of $\langle n^2(t)\rangle$ versus time graph (see section 7 of SI). $D_n$ obtained from the equilibrium simulation is used to estimate the water flux under different pressure gradients using Eq. \ref{3}. The collective diffusion model is able to predict accurately the water fluxes calculated from the non-equilibrium pressure gradient simulation using only equilibrium simulation data.\par

We have also performed equilibrium and non-equilibrium simulations of CNT membrane having chirality (9,9) under three different pressure gradients 50 MPa, 200 MPa, 400 MPa, to have a direct comparison with the PNT. All the CNTs have hydrogen as the end-functionalization group. The flux values obtained for (9,9) CNT from the non-equilibrium simulations and the collective diffusion model are given in the SI table S12. The collective diffusion model accurately predicts the flux data over a wide range of pressures, signifying the robustness of the model. We find that the water flux of (9,9) CNT is higher than that of the (9,9) PNT(H). The latter can be explained by comparing the free energy profiles (PMFs) for water transport through CNT and PNT (see Fig. 2). The PMF of water is very smooth for CNT, whereas it is rugged for PNT. So, water gets trapped transiently in local minima, which reduces the permeability of PNT(H) compared to CNT of the same chirality. 
Note that the ion rejection \% of (9,9) CNT is much lower (27\% at 200 MPa) than that of (9,9) PNT (almost 100\% up to 200 MPa). This is the main advantage of PNT that it can reject salts with a reasonably high water permeability, which thus makes it one of the potential membrane for water filtration. 
\par

\begin{figure*}[htbp]
 \centering
 \includegraphics[width=1.0\linewidth,keepaspectratio=true]{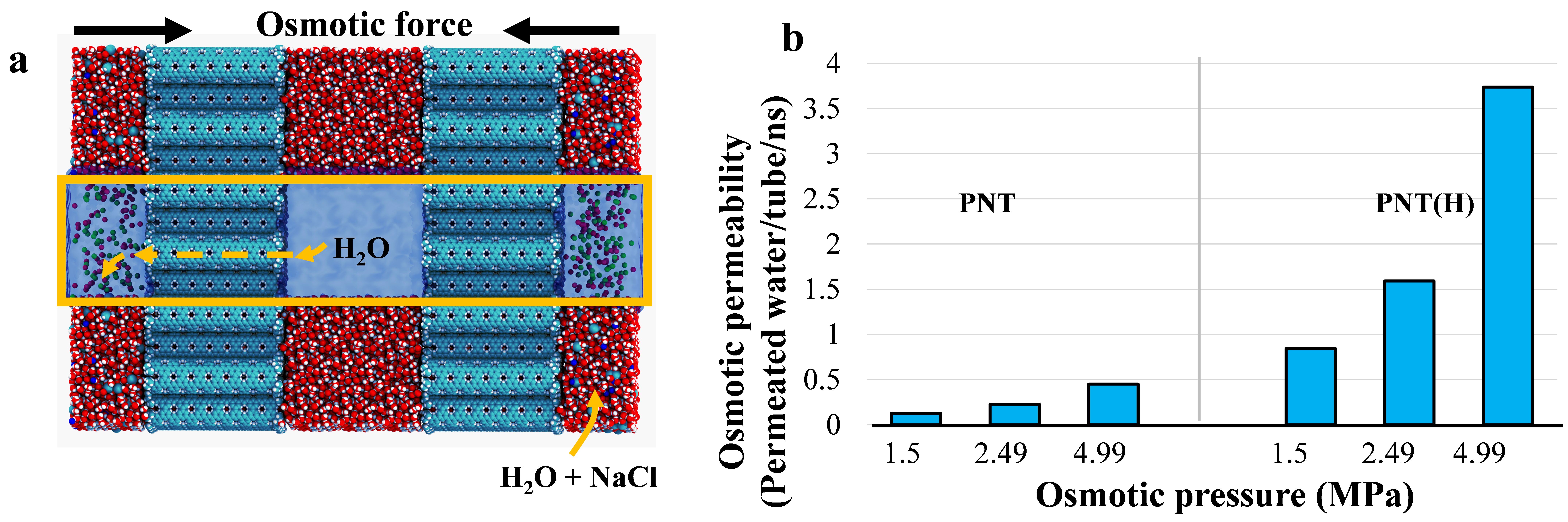}
 \caption{Osmotic permeation of water across the PNT membrane. (a) Initial setup of the system for the osmotic permeability calculation. The two PNT membranes are separated by two water compartments, one filled with pure water and the other with the saltwater of salinity 600 mM, 1000 mM, or 2000 mM. The rectangular box marked by yellow line represents the simulation unit shell. (b) Osmotic permeability as a function of osmotic pressure ($\Pi = CRT$, C=salt concentration, R=universal gas constant, T= temperature) for PNT and PNT(H) membranes.}
 \label{fig4}
\end{figure*}

The osmotic permeation across a membrane due to a salt concentration gradient is one of the crucial thermodynamic quantity in the context of RO membrane-based water desalination\cite{31}. To introduce an osmotic pressure gradient in the simulation, we assemble two PNT membranes separated by two water compartments, one filled with pure water and the other with the saltwater of salinity 600 mM, 1000 mM, or 2000 mM which corresponds to the osmotic pressure 1.5 Mpa, 2.49 MPa, or 4.99 MPa, respectively (Fig. \ref{fig4}a). A similar simulation set-up was used earlier by Kalra et al.\cite{30} to study the osmotic transport of water across CNT membranes. The sub-nanometer pores of the PNT membrane allow passage of water but not Na$^+$ and Cl$^-$ ions; thus, an osmotic pressure gradient across the membrane is created. As a result, during the simulation, the pure water compartment drains out, and consequently, the saline water compartment expands. This water flow pushes the two PNT membranes in the opposite direction, which eventually leads to the association of the two PNT membranes in the energy minimized configuration (see Fig. S8 of SI). This simulation set-up allows us to study the collective diffusion of water due to an osmotic pressure gradient\cite{31}.\par
To calculate the osmotic permeability, we first estimate the water permeation across the two membranes. Then, from the linear region of the average water permeation versus time data, we calculate the slope, and consecutively the net osmotic permeability. The calculated the osmotic permeability across the PNT membrane is increases linearly with the increase in the osmotic pressure (Fig. \ref{fig4}b). Moreover, the permeability of PNT(H) membrane from this simulation setup is coming around 3.73 water molecules/ns/nanotube at an osmotic pressure of 4.99 MPa, whereas, the permeability is 3.16 water molecules/ns/nanotube at 5 MPa in case of non-equilibrium hydrostatic pressure-gradient simulation. As the osmotic permeability of the PNT membrane is much higher than conventional RO-based plants\cite{31}, the PNT membrane is an excellent nanomaterial that can be used for energy-efficient water desalination. The described method can be used to compute the osmotic permeability of any membrane directly from equilibrium MD simulation.\par
\section{Conclusion}
In summary, we have characterized the structure, thermodynamics, and kinetic behavior of water and ions  transport through PNT, which has been recently synthesized in experiments\cite{28}. Our results demonstrate the fascinating properties of PNT that can be utilized for a variety of applications, from water desalination to selective bio-mimicking ion channels. Employing non-equilibrium pressure-driven water flow simulations, we have shown that the PNT membrane can reject all ions while allowing faster water flows with permeabilities several orders of magnitude higher than many of the conventional membranes for water filtration reported in the literature. The collective diffusion model is shown to quantitatively reproduce, over a wide range of pressure gradients,  water flow rates obtained from simulations. Our proposed osmotic permeability calculation method from equilibrium MD simulation shows that the osmotic permeability across these membranes is very high and agrees well with the non-equilibrium hydrostatic pressure gradient simulation permeability. Moreover, our proposed method for the osmotic permeability calculation is straightforward and can be applied to other membrane-based systems. We hope that our results will encourage experimentalists to assess the PNT membrane for water filtration and will aid in designing next-generation membranes for water desalination technology. \par

\section{Method}
\subsection{Model}
The atomic model of the CNT is built using Visual molecular dynamics (VMD)\cite{humphrey1996vmd} software and then using an in-house python script, the PNT is constructed by removing specific carbon atoms from the CNT. Two different nanotubes of chirality (9,9) and (12,12) and each of length 50 {\AA} is built. Using the XLEAP module of Ambertools\cite{duke2016ambertools}, hydrogen atoms on specific carbon atoms are added. To see the effect of the end functional group on desalination, we simulate PNT with two different end groups, the one with an experimentally synthesized tertiary butyl (t-Bu) group (referred to as PNT) and the other is with hydrogen (referred to as PNT(H)) (Fig. \ref{fig1}). The end groups (t-Bu or hydrogen) are added on the PNT using the Gaussian\cite{frisch2009gaussian} software. Each nanotube is solvated in TIP3P\cite{jorgensen1983comparison} water with a water buffer of more than 20 {\AA} in length in all three directions. Desired number of ions (NaCl or KCl) are added to the solvated system to achieve the seawater salinity i.e., 600mM. Further details of the simulated systems are given in the next section.\par
\subsection{Calculation of partial atomic charges}
\begin{figure*}[htbp]
 \centering
 \includegraphics[width=1.0\linewidth,keepaspectratio=true]{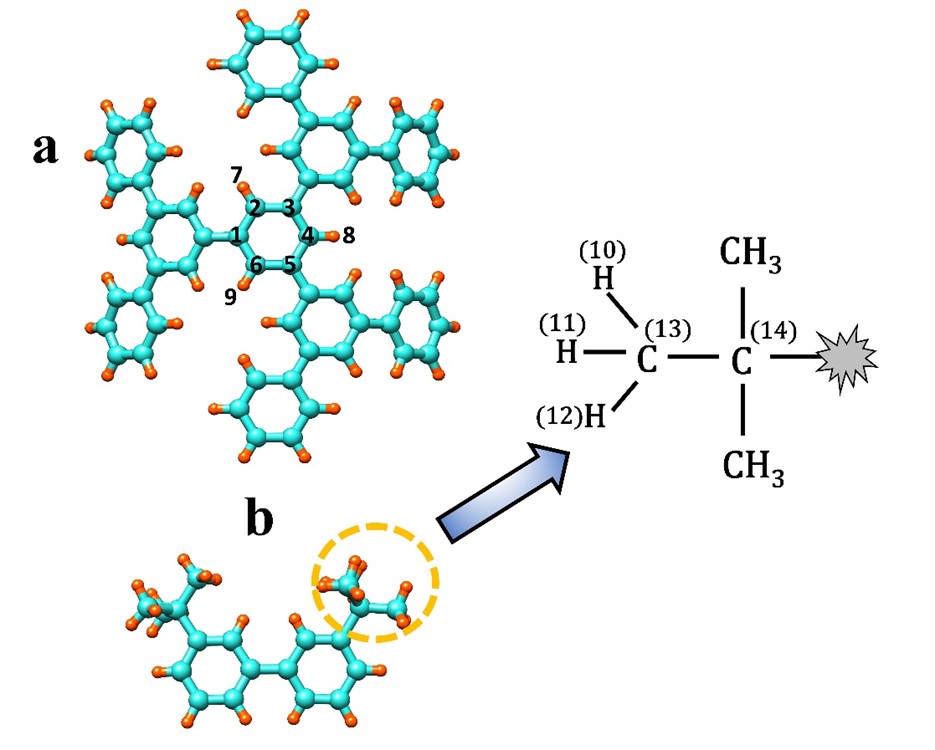}
 \caption{(a) The segment of the PNT molecule utilized for RESP charge calculation. The charge of the middle segment (the numbered particles) is used in our simulation. (b) The fragment used to calculate the charge on the tertiary butyl group. The molecule enclosed by a circle is the  tertiary butyl group. (c) Chemical structure of tertiary butyl group. The numbers are used for indexing to provide the partial charges. The RESP charges are provided in Table 1.}
 \label{fig5}
 \end{figure*}

To evaluate the partial charges on the PNT atoms, a specific segment of the molecule is selected. This is done in such a way that the middle benzene unit has identical surroundings to the primordial PNT molecule [Fig. \ref{fig5}]. Then to determine the electrostatic potential (ESP) of that segment we employ quantum mechanical calculation using HF/6-31G* basis set with the Gaussian software package \cite{frisch2009gaussian}. The HF/6-31G* level of theory is widely used for charge calculation compatible with AMBER force fields. The restraint electrostatic potential (RESP) charges are then derived from the ESP, following the same approach as used in the origin RESP paper by Kollman et al. \cite{bayly1993well,wang2000well}. We take the RESP charges on the atoms of the middle benzene unit for all the simulations. To calculate the partial atomic charges on the end t-Bu group, we choose a different chunk of the molecule containing two benzene rings and two t-Bu groups [Fig. \ref{fig5}b]. The partial atomic charges of the end t-Bu groups are estimated following the same procedure as mentioned above with restraining the charges on the benzene group as evaluated before. The partial charges on the PNT atoms remain fixed during the simulation. All the charges are enlisted in Table 1.
\par
\subsection{Force-field parameters}
The bonded and non-bonded interactions for CNT carbon atoms are modeled using the second generation generalized amber force field (GAFF2) \cite{cornell1995second}. We take Joung/Chetham Lennard-Jones parameters \cite{joung2008determination} to describe the interaction of ions with water and CNT or PNT. To determine the bonded and non-bonded parameters of PNT, we use the Antechamber module \cite{wang2001antechamber} with parameters taken from GAFF2.\par

\begin{table*}
\centering
	\caption{Values of the calculated charges.\\\hspace{\textwidth}
	The index number of the atoms are given in Fig. \ref{fig5}}
\label{table}
\begin{center}
\resizebox{\textwidth}{!}{\begin{tabular}{|cccccc|}
\hline
	Index number	&\multicolumn{1}{|c|}{Atom type}&\multicolumn{1}{c|}{RESP charge} &\multicolumn{1}{||c|}{Index number}	&\multicolumn{1}{|c|}{Atom type}&\multicolumn{1}{c|}{RESP charge}\\
\hline
	1	&\multicolumn{1}{|c|}{C}&\multicolumn{1}{c|}{-0.0304} &\multicolumn{1}{||c|}{8}	&\multicolumn{1}{|c|}{H}&\multicolumn{1}{c|}{0.1374}\\
\hline
	2	&\multicolumn{1}{|c|}{C}&\multicolumn{1}{c|}{-0.1070} &\multicolumn{1}{||c|}{9}	&\multicolumn{1}{|c|}{H}&\multicolumn{1}{c|}{0.1374}\\
\hline
	3	&\multicolumn{1}{|c|}{C}&\multicolumn{1}{c|}{-0.0304} &\multicolumn{1}{||c|}{10}	&\multicolumn{1}{|c|}{H}&\multicolumn{1}{c|}{0.17385}\\
\hline
	4	&\multicolumn{1}{|c|}{C}&\multicolumn{1}{c|}{-0.1070} &\multicolumn{1}{||c|}{11}	&\multicolumn{1}{|c|}{H}&\multicolumn{1}{c|}{0.1739}\\
\hline
	5	&\multicolumn{1}{|c|}{C}&\multicolumn{1}{c|}{-0.0304} &\multicolumn{1}{||c|}{12}	&\multicolumn{1}{|c|}{H}&\multicolumn{1}{c|}{0.1739}\\
\hline
	6	&\multicolumn{1}{|c|}{C}&\multicolumn{1}{c|}{-0.1070} &\multicolumn{1}{||c|}{13}	&\multicolumn{1}{|c|}{C}&\multicolumn{1}{c|}{-0.5469}\\
\hline
	7	&\multicolumn{1}{|c|}{C}&\multicolumn{1}{c|}{0.1374} &\multicolumn{1}{||c|}{14}	&\multicolumn{1}{|c|}{C}&\multicolumn{1}{c|}{0.0756}\\
\hline
\end{tabular}}%

\end{center}

\end{table*}
For the tertiary butyl group, all three CH$_3$ groups have the same value of RESP charge. The partial atomic charges, modified force-field and the built structures are available upon request. \par
\subsection{Equilibrium MD simulations}
The built structures are first subjected to several steps of energy minimization to eliminate any bad contacts between the atoms. This involves 2000 steps of steepest descent and 3000 steps of conjugate-gradient minimization. The nanotube is restrained to its initial position with a harmonic force constant of 500 kcal/mol.{\AA$^{-2}$} during the minimization. Then the systems are subjected to another 5000 steps of minimization with a reduced harmonic force constant of 20 kcal/mol.{\AA$^{-2}$} on the nanotube. Then the restraint on the nanotube is gradually decreased from 20 kcal/mol.{\AA$^{-2}$} to 0 kcal/mol.{\AA$^{-2}$} with steps of 5 kcal/mol.{\AA$^{-2}$} in 20000 energy minimization steps. \par
The energy minimized structures are then gradually heated from 10 K to 300 K within 50 ps in the NPT ensemble. During the heating, the nanotube is restrained to its initial position with a harmonic force constant of 20 kcal/mol.{\AA$^{-2}$}. We then equilibrate the systems by performing 2 ns simulation in the NPT ensemble at T = 300 K and P = 1 bar. The equilibrated structures are then subjected to more than 100 ns long MD simulation in the NVT ensemble. The temperature regulation is achieved by Langevin thermostat\cite{van1988leap} with a collision frequency of 2 ps$^{-1}$. Unless specified, we use the Berendsen weak coupling method \cite{berendsen1984molecular} with isotropic pressure scaling to maintain the pressure during NPT simulations. Shake algorithm \cite{ryckaert1977numerical} is employed to constrain the hydrogen atoms during the simulation which allows us to use an integration time step of 2 fs. Long-range Columbic interactions are evaluated by implementing particle mesh Ewald \cite{darden1993particle} method with a  real-space cutoff of 10 {\AA}. Similar methodologies have been successfully implemented in several of our previous studies \cite{moid2019microscopic, sahoo2018translocation}.\par

\subsection{Free-energy and electrostatic potential map calculations}
The potential of mean force (PMF) along the long axis of a nanotube (taken as the reaction coordinate, z, here) are obtained employing the umbrella sampling (US) method\cite{torrie1977nonphysical}. We use a harmonic biasing potential of the form k$_z$ [(z-z$_0$)]$^2$  to restrain a molecule (water, [Na]$^+$, or [Cl]$^-$) at a position (z$_0$) along z with a force constant of 1 kcal/mol.{\AA$^{-2}$}. An additional restraint of very low force constant of 0.028 kcal/mol.{\AA$^{-2}$} in both X- and Y- direction is added when the molecule is outside the nanotube to restrain the molecules in the circular region of area equivalent to the nanotube pore area. A similar methodology has been implemented to calculate the free energy of ions inside the CNT pore by Corry et. al.\cite{11,12} We use 1 {\AA} bin window size to sample the whole range of the reaction coordinate. In each window, we run 4 ns of NVT simulation to accumulate the trajectory. The PMF profile is then constructed employing the weighted histogram analysis method \cite{kumar1992weighted}. We calculate PMF from the center of the nanotube to the end of the box and then by symmetry we constructed the whole PMF profile. \par
The electrostatic potential, $\phi(r)$  has been computed by solving the Poisson equation,
\begin{equation}
    \nabla^{2} \phi(r) = -4\pi \sum_{i} \rho_{i}(r)
\end{equation}
where the sum, $i$ runs over all the atoms and the charge density, $\rho_i(r)$ is approximated by a spherical Gaussian
\begin{equation}
    \rho_{i}(r) = q_i \left(\frac{\omega}{\sqrt{\pi}}\right) e^{-\omega^{2}|r-r_i|^2}
\end{equation}
Here $\omega$ is the inverse width of the Gaussian as defined by Aksimentiev $et$ $al.$ in the original paper of electrostatic potential map calculation\cite{aksimentiev2005imaging}. The instantaneous potential profile, $\phi(r)$, is then averaged over the last 25 ns of the whole simulation and considering all charged atoms present in system.\par

\subsection{Membrane modeling and hydrostatic pressure gradient simulation}
A porous membrane is created by arranging the twelve (9,9) PNT molecules, each of length 50 {\AA}, in a hexagonal lattice (see Figure \ref{fig3}a). This is done to achieve the closest packing structure possible for the PNT membrane. Also, carbon nanotube membranes are known to exist in the hexagonal close packing structure\cite{fornasiero2008ion}, which prompted us to build the membranes in the hexagonal lattice. Another possible structure of the PNT membrane can be tetragonal as the crystal structure of the membrane provided in the original paper was tetragonal, but many fullerene (C60) molecules were enclosed in between different parts of the membrane\cite{sun2019finite}. Furthermore, it is not yet known that whether the PNT molecules will form the tetragonal lattice without such bulky fullerene molecules. So we decided to carry out our study of desalination on hexagonally packed membrane. Periodic boundary conditions are implemented such that it forms a continuous 2D membrane [Figure \ref{fig3}a and b]. A water layer of 30 {\AA} on either side of the membrane is added with an appropriate number of Na$^+$ and Cl$^-$ placed randomly such that the net concentration of NaCl in water becomes 600 mM. To check the thermodynamic stability of the assembled membrane, we then carry out 100 ns of NPT simulation with anisotropic pressure coupling in the $z$-direction. Final structures from the NPT simulation are picked for the hydrostatic pressure difference simulations. The hydrostatic pressure is implemented across the membrane by a method originally developed by Zhu $et$ $al.$ \cite{zhu2002pressure} A constant force f is applied to all the water molecules and ions along the $z$-direction. Due to the periodic nature of the box, this induces a pressure gradient, $\Delta P$, across the membrane given by
\begin{equation}
    \Delta P = -\frac{Nf}{A}
\end{equation}
where N is the total number of molecules and A is the cross-sectional area. The minus sign arises because of the directionality of force is opposite to that of pressure gradient. Six different simulations of hydrostatic pressure gradient of 5, 50, 100, 200, 300, 400 MPa are carried out for 100 ns each, in the NVT ensemble. Note that the choice of simulation thermostat can have a big influence on the results\cite{thomas2015thermostat}. Every thermostat has its own advantages and disadvantages. Langevin thermostat is dynamically stable and reproduces a canonical ensemble over long timescales, but the frictional and random forces can perturb the transport properties. Langevin thermostat focuses on replicating energetics and maintaining dynamic stability. Since the free-energy and electrostatics calculation requires a thermostat which replicates correct energetics, we choose Langevin. Later, we did not modify the thermostat for the uniformity of the study. The use of a Langevin thermostat for mass transport may not be ideal as it decreases coupled motion of atoms and reduces flow rates. \par
\subsection{Osmosis}
In order to introduce an osmotic pressure gradient, we assemble two PNT membranes separated by two water compartments, one filled with pure water and the other with the saltwater of salinity 600 mM, 1000 mM, or 2000 mM. All the built systems with different NaCl salinities are then subjected to 300 ns to 500 ns MD simulation in the NPT ensemble. Berendsen barostat with an anisotropic pressure coupling and Langevin thermostat is used for these simulations. For the osmotic permeability calculation, the linear region of the time evolution of water flux, as shown in Figure S7, is used. 

\subsection{Softwares used for simulations, visualizations, and analyses}
All the equilibrium MD simulations are performed using PMEMD and PMEMD.CUDA\cite{gotz2012routine, salomon2013routine} modules of AMBER package. The non-equilibrium pressure gradient MD simulations are performed in NAMD2\cite{phillips2005scalable,kale1999namd2} software package. Visualizations are done using VMD\cite{humphrey1996vmd} and UCSF Chimera software\cite{pettersen2004ucsf}. Analyses are performed using CPPTRAJ\cite{roe2013ptraj}, MDTRAJ\cite{mcgibbon2015mdtraj}, VMD\cite{humphrey1996vmd}, python and TCL scripts.

\section{Supporting information}
Methods, Details of the simulated systems, Validation of the partial atomic charges and force-field, Density of water in the vicinity of the nanotube, Structure of ions and water molecules inside different nanotubes, Collective diffusion model, Osmotic pressure calculation.
Supporting information available online.

\section{Acknowledgements}
A.K.S. and M.M. contributed equally to this work. We thank computational support through TUE-CMS, IISc. S.N. and M.M. acknowledge SRF fellowship from CSIR, India. A.K.S. acknowledges IISc for the institute RA fellowship. We thank Prof. Manish Kumar for useful comments and suggestions on the manuscript. 

\bibliography{main}
\end{document}